\documentclass[12pt,preprint]{aastex}

\newcommand{\fermi}{{\it Fermi}-LAT }
\newcommand{\gppr}{\stackrel{>}{\scriptstyle \sim}}
\newcommand{\lppr}{\stackrel{<}{\scriptstyle \sim}}

\slugcomment{ApJ submitted}

\shorttitle{Evidence for second high-energy component in Cen A}
\shortauthors{Sahakyan et al.}

\begin{document}


\title{Evidence for a second component in the high-energy core emission from Centaurus A?}


\author{N. Sahakyan\altaffilmark{1}}
\affil{ICRANet, Piazz della Repubblica 10, I-65122 Pescara, Italy\\
Institute for Physical Research, NAS of Armenia, Ashtarak-2, 0203, Armenia}

\author{R. Yang}
\affil{Key Laboratory of Dark Matter and Space Astronomy, Purple Mountain Observatory, CAS, Nanjing, 210008, China}

\author{F.A. Aharonian}
\affil{Dublin Institute for Advanced Studies, 31 Fitzwilliam Place, Dublin 2, Ireland}

\and

\author{F.M. Rieger}
\affil{Max-Planck-Institut f{\"u}r Kernphysik, P.O. Box 103980, 69029 Heidelberg, Germany}




\begin{abstract}
We report on an analysis of \fermi data from four year of observations of the nearby radio galaxy Centaurus A
(Cen~A). The increased photon statistics results in a detection of high-energy ($>\:100$ MeV) $\gamma$-rays up to 50
GeV from the core of Cen~A, with a detection significance of about 44$\sigma$. The average gamma-ray spectrum of the
core reveals evidence for a possible deviation from a simple power-law. A likelihood analysis with a broken power-law
model shows that the photon index becomes harder above $E_b \simeq 4$ GeV, changing from $\Gamma_1=2.74\pm0.03$
below to $\Gamma_2=2.09\pm0.20$ above. This hardening could be caused by the contribution of an additional high-energy
component beyond the common synchrotron-self Compton jet emission. A variability analysis of the light curve with
15-, 30-, and 60-day bins does not provide evidence for variability for any of the components. Indications for a possible
variability of the observed flux are found on 45-day time scale, but the statistics do not allow us to make a definite conclusion
in this regards. We compare our results with the spectrum reported by H.E.S.S. in the TeV energy range and discuss possible
origins for the hardening observed.
\end{abstract}





\section{Introduction}
The prominent radio galaxy Centaurus A (NGC 5128), at a distance of $\simeq 3.8$ Mpc ($1' \simeq 1.1$ kpc) \cite{harris10},
is the closest active galaxy to Earth. Often regarded as a prototype Fanaroff-Riley Class I \cite{fanaroff74} radio source
and as a misaligned BL Lac-type object at higher energies \cite{morg,chiaberge01}, its proximity has made it one of the
best-studied extragalactic objects over a wide range of frequencies \cite[e.g.][for review]{isr98}. Cen A is known
for a complex and extended radio morphology, with two giant outer lobes extending over $\sim10^\circ$ and oriented
primarily in the north-south direction \cite{feain11}. Optical images reveal the bright host galaxy bulge ($\sim5'$ bulge
radius) and the famous, warped dark lane of gas, dust and young stars ($\sim 12' $ in east-west extension) which
obscures the inner part of the galaxy. Chandra X-ray observations show a one-sided, kpc-scale (up to $\sim4.5$ kpc in
projection) jet composed of several bright knots and diffuse emission, while high-resolution radio VLBI observations
have also resolved jet and counter-jet features on sub-parsec scales into discrete components \cite{kraft02,mueller11}.
These and related observations suggest that Cen A is a non-blazar source with its jet inclined at a rather large viewing
angle $\gppr 50^\circ$ and characterized by moderate (radio) bulk flow speed $\lppr 0.5c$ \cite{tingay98,hardcastle03,mueller11}.
Its bolometric luminosity of $L\sim10^{42}$ erg/s \cite{meisenheimer07,wolk10}, accompanied by indications for the lack
of a dust torus, is thought to be powered by gas accretion onto a supermassive black hole of mass $M_{\rm BH}\simeq
(3-12) \times 10^7 M_\odot$ \cite{marconi06,cappellari09}.

At MeV energies, Cen A has been observed with both OSSE (0.05-4 MeV) and COMPTEL (0.75-30 MeV) onboard the
Compton Gamma-Ray Observatory (CGRO) in the period 1991-1995 \cite{sten98}. An agreement of the OSSE spectrum
with the COMPTEL one in the transition region around 1 MeV, and correlated variability has been found \cite{sten98}. At
higher energies, a marginal ($3 \sigma$) detection of gamma-rays from the core of Cen A was reported with EGRET
(0.1-1.0 GeV), but due its large angular resolution the association with the core remained rather uncertain \cite{har99}.
Unlike the initial variability (month-type?) seen at lower energies, the flux detected by EGRET appeared stable during the
whole period of CGRO observation \cite{sree99}.

At VHE ($>100$ GeV) energies, Cen A has also been detected (with a significance of $5\sigma$) by the H.E.S.S. array
based on observations in 2004-2008. The results show an average VHE spectrum compatible with a power law of photon
index $\Gamma=2.73\pm0.45_{stat}\pm0.2_{syst}$ and an integral flux $F(E > 250\mathrm{\,GeV}) = (1.56 \pm 0.67_{\mathrm{stat}})
\times 10^{-12}$\,cm$^{-2}$\,s$^{-1}$\cite{aharonian09}. No evidence for variability has been found in the H.E.S.S. data set,
but given the weak signal no certain conclusions can be drawn.

\fermi has reported the detection of high energy (HE, $>100$ MeV) gamma-rays from both the core (i.e., within
$0.1^\circ$) and the giant radio lobes \cite{abdo10b,abdo10a}. The analysis of 10 months of data revealed a point-like HE
emission region coincident with the position of the radio core of Cen A, the emission being well described by a power-law
function with a photon index $\approx2.7$, quite similar to the one in the VHE regime. Also, no variability (on 15 d and 30 d time
scales) has been found. A simple extrapolation of the HE power-law spectrum to the VHE regime, however, fails to account
for the TeV core flux as measured by H.E.S.S., which could indicate the need for an additional contribution towards the highest
energies.
The giant lobes were detected with a significance of 5$\sigma$ and 8$\sigma$ for the northern and the southern structure,
respectively \cite{abdo10a}. The HE lobe spectrum could be described by a power-law function extending up to 2 or 3 GeV
with photon indices of $\Gamma\approx2.6$. A recent analysis of a three times larger data set has confirmed the existence
of these HE lobes, but also showed that the HE emission extends well beyond the WMAP radio image \cite{yang}.

The apparent lack of significant variability features at GeV and TeV energies has so far precluded robust inferences as to
the physical origin of the core emission in Cen A. Unfortunately, the resolutions of current gamma-ray instruments is not
sufficient to localize the gamma-ray emitting region(s) either: The angular resolutions of both the H.E.S.S. array ( $\sim0.1^\circ$)
and \fermi ($0.1^\circ$ -$1^\circ$, depending on energy) correspond to linear sizes of the gamma-ray emitting region(s) of
about 5 kpc or larger. This $\sim5$-kpc-region contains several potential gamma-ray emitting sites such as the central black
hole, the sub-pc- or the kpc-scale jet etc. Based on the reported results, one thus cannot distinguish whether the gamma-rays
observed from the core in Cen A originate in compact or extended regions. This motivated us to have a new look on the core
emission based on four year of \fermi data.
\section{Fermi-LAT Data Analysis}\label{sec2}
\subsection{Data Extraction}\label{sec2.1}
\fermi on board the Fermi satellite is a pair-conversion telescope designed to detect high-energy $\gamma$-rays in the
energy range 20 MeV - 300 GeV \cite{atwood09}. It constantly scans the entire sky every three hours and is always
in survey mode.

For the present analysis we use publicly available \fermi $\sim4$~yr data from 4th August 2008 to 1st October 2012
(MET 239557417--370742403). We use the Pass 7 data and analyze them using the Fermi Science Tools v9r27p1
software package. The entire data set was filtered with {\it gtselect} and {\it gtmktime} tools and retained only events
belonging to the class 2, as is recommended by the Fermi/LAT science team\footnote{http://fermi.gsfc.nasa.gov/ssc/data/analysis/documentation/\newline Cicerone/Cicerone\_Data\_Exploration/Data\_preparation.html}.
To reject atmospheric gamma-rays from the Earth's limb, events with zenith angle $<100$ deg are selected. The standard
binned maximum likelihood analysis is performed using events in the energy range 0.1--100\, GeV extracted from a
$10^{\circ}$ region centered on the location of Cen A, which is referred to as 'region of interest' (ROI). The fitting model
includes diffuse emission components and gamma-ray sources within ROI which are not associated with Cen A (the
model file is created based on \fermi second catalog \cite{abdo11}. In the model file, the giant radio lobes were modeled
using templates from WMAP-k band observation of the source which is extracted from NASA's SkyView. Although our
previous results showed that the HE extension and the radio lobe regions do not perfectly match \cite{yang}, this does
not affect the central parts of relevance here. The background was parameterized with the files
gal\textunderscore 2yearp7v6\textunderscore v0.fits and iso\textunderscore p7v6source.txt and the normalizations of
both components were allowed to vary freely during the spectral point fitting.
\subsection{Spectral Analysis}\label{sec2.2}
\begin{figure}[thb]
   \centering
   \plotone{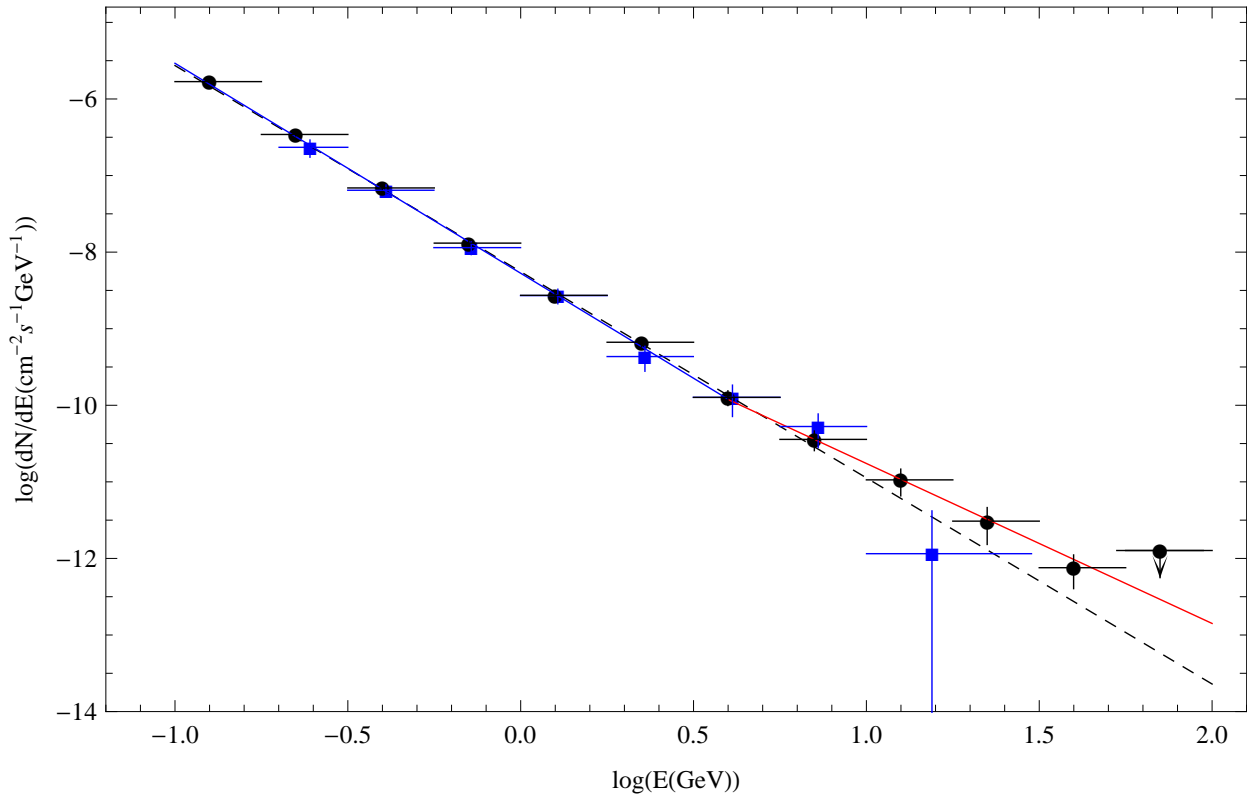}
   \caption{Average high-energy gamma-ray ($>$100 MeV) spectrum of the core of Cen A (black points - this work) as
   compared to the one based on the initial 10 month data set \citep[blue squares - ][]{abdo10b}. The dashed black line shows the
   power-law function determined from the gtlike. The blue and the red line show power-law fits to the energy bands below and
   above $E_b \simeq 4$ GeV, respectively.}
    \label{fg1}
\end{figure}

Initially the continuum gamma-ray emission of the core of Cen A is modeled with a single power law. The normalization and
power-law index are considered as free parameters then the binned likelihood analysis is performed. From a binned {\it gtlike}
analysis, the best-fit power-law parameters for the core of Cen A are
\begin{equation}
\left(\frac{dN}{dE}\right)_{P}=(2.73\pm0.12)\times10^{-9}\left(\frac{E}{100\;\mathrm{MeV}}\right)^{-2.69\pm0.03}\,.
\label{pl}
\end{equation}
This corresponds to an integral flux of
\begin{equation}
F_{\gamma}=(1.61\pm0.06)\times10^{-7}\:\mathrm{photon\:cm}^{-2}s^{-1},
\label{pl1}
\end{equation}
with only statistical errors taken into account. The test statistic (defined as TS = 2(log $L$ - log $L_0$), where $L$ and $L_0$
are the likelihoods when the source is included or not) is $TS$ = 1978 above 100 MeV, corresponding to a $\approx 44\: \sigma$
detection significance. The results are consistent with the parameters found in \cite{abdo10b}, namely photon index $\Gamma
=2.67\pm0.08$ (between 200 MeV and 30 GeV) and integral flux $(1.50\pm0.37)\times10^{-7}$ ph cm$^{-2}$s$^{-1}$ above $100$ MeV (model B).
Figure~\ref{fg1} shows the spectrum of the core of Cen A obtained by separately running {\it gtlike} for 12 energy
bands, where the dashed line shows the best-fit power-law function for the data given in Eq.~(\ref{pl}). For the highest energy bin
(56.2-100 GeV), an upper limit is shown.
The spectrum shows a tendency for a deviation from a single power-law model with respect to the data above several GeV.
Indeed, a $\chi^{2}$ fit of the power-law model to the data gives a relatively poor fit with $\chi^{2}=39.7$ for 9 degrees of
freedom (dof), and its probability is $P(\chi^{2})<2\times10^{-5}$. In order to investigate this in more detail, the core spectrum
is modeled with a broken power-law model and {\it gtlike} tool is  retried. The best-fit broken power-law parameters are
\begin{eqnarray}
\left(\frac{dN}{dE}\right)_{BP}=(1.19\pm0.08)\times10^{-13}\left(\frac{E}{E_{b}}\right)^{-\Gamma_{1,2}},
\label{pl2}
\end{eqnarray}
and
\begin{equation}
\centering
F_{\gamma}=(1.67\pm0.06)\times10^{-7}\:\mathrm{photon\:cm}^{-2}s^{-1},
\label{pl3}
\end{equation}
with $\Gamma_1=2.74\pm0.02$ and $\Gamma_2=2.12\pm0.14$ below and above $E_b=(4.00\pm0.09)$ GeV, respectively. In order
to compare the power-law and the broken-power-law model, a log likelihood ratio test between the models is applied. The test statistic is
twice the difference in these log-likelihoods, which gives 9 for this case. Note that the probability distribution of the test statistic can be
approximated by a $\chi^{2}$ distribution with 2 dof, corresponding to different degrees of freedom between the two functions. The results
give $P(\chi^{2})=0.011$, which again indicates a deviation from a simple power-law function.
The results of the data analysis with a broken power-law model reveal a hardening of the (average) gamma-ray core spectrum towards
higher energies. The "unusual" break at 4 GeV could most naturally be explained by a superposition of different spectral components.
In order to study this deeper, we divide the data set into two parts, i.e., (0.1- 4) GeV and (4-100) GeV. (Note that the 4 GeV-value is obtained
from binned maximum likelihood analyses). The core spectrum of Cen A in both energy ranges is then modeled with a power-law function
and the {\it gtlike} tool is separately applied to these two energy bands. The photon index and flux between 100 MeV and 4 GeV are
$\Gamma_1=2.74\pm0.02$ and $F_{\gamma}=(1.68\pm0.04)\times10^{-7}$ photon cm$^{-2}$s$^{-1}$, respectively, and the test
statistics gives TS=1944. The result is shown with a blue line in Fig.~\ref{fg1}. On the other hand, for the energy range (4-100) GeV we
obtain $\Gamma_2=2.09\pm0.2$ and $F_{\gamma}=(4.20\pm0.64)\times10^{-10}$ photon cm$^{-2}$s$^{-1}$, respectively, and a TS
value of 124.4, corresponding to a $\approx 11\: \sigma$ detection significance. This component is depicted with a red line in Fig. \ref{fg1}.
\section{Temporal Variability}\label{sec3}
Variability, if present, could provide important constraints on the emitting region(s). An observed HE flux variation on time scale $t_{\rm var}$,
for example, would limit the (intrinsic) size of the gamma-ray production region to $R'\leq \frac{\delta_{D}}{1+z} c t_{\rm var}$. However,
previous HE and VHE gamma-ray observations of the core of Cen A with \fermi \cite{abdo10b} and H.E.S.S. \cite{aharonian09} did not find
evidence for significant variability. Here we investigate whether the longer (4~yr) data set employed changes this situation. We thus divide
the whole data set (from August 4th 2008 to October 1st 2012) into different time bins and generate light curves using the unbinned likelihood
analysis with {\it gtlike}. Due to limited photon statistics the shortest time scale that one can probe is 15 days. In our analysis we generate light
curves in 15, 30, 45 and 60 day bins. The normalization of the core and background point sources are treated as free parameters, but the
photon indices of all sources and the normalization of the lobes are fixed to the values obtained in 100 MeV-100 GeV energy range for the
whole time period. Since no variability is expected for the underlying background diffuse emission, the normalization of both background
components is fixed to the values obtained for the whole time period.
\begin{figure*}[thb]
   \centering
    \plotone{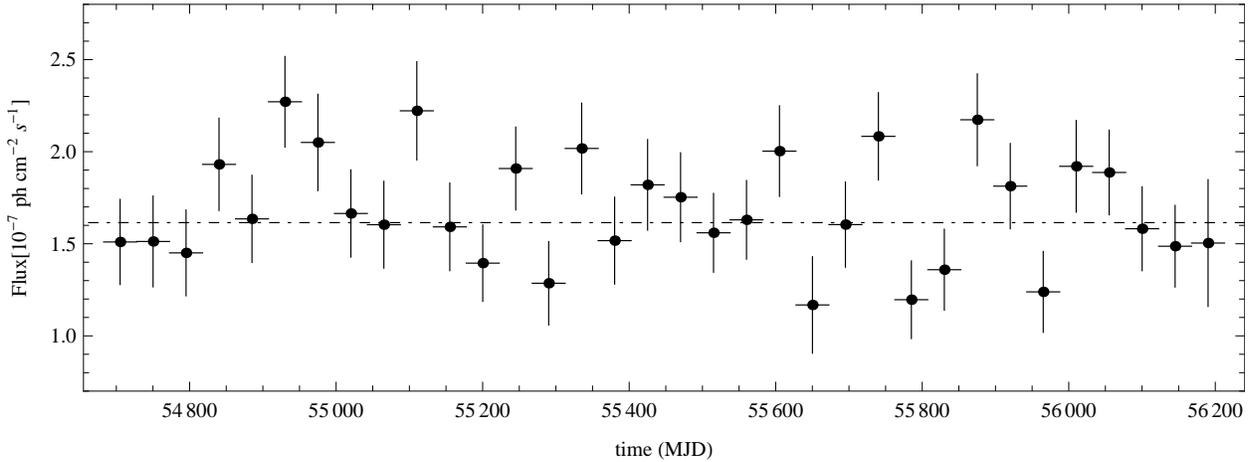}
    \caption{Gamma-ray light curve from August 4th 2008 to October 1st 2012. The bin size is 45 day. The background diffuse emission
    (both galactic and extragalactic) is fixed to the best-fit parameters obtained for the overall time fit. While some variability may be
    present, limited statistics do not yet allow to make definite conclusions.}%
    \label{fg2}
\end{figure*}
To search for variability, a $\chi^{2}$ test was performed. The result for the light curve with 15 day bins is $\chi^{2}/d.o.f.=1.22$ and the
probability is $P(\chi^{2})=0.07$. For the light curves with 30 day and 60 day bins we find $\chi^{2}/d.o.f.=1.37$ and $\chi^{2}/d.o.f.=1.32$,
corresponding to $P(\chi^{2})=0.04$ and $P(\chi^{2})=0.127$, respectively. These results are consistent with no variability. Interestingly
however, a similar test for the light curve with 45 day bins gives in $\chi^{2}/d.o.f.\approx1.61$ and $P(\chi^{2})=0.015$, indicating a possible
variability on 45-day time scale. Unfortunately, because of limited statistics, we cannot make a definite conclusion in this regard. The light
curve with 45 day bins is shown in Fig.~\ref{fg2}, with the dot-dashed line indicating the flux from the source for the whole time period
(result of likelihood analysis).
\section{Discussion and Conclusion}\label{sec4}
In the case of high-frequency-peaked BL Lac objects, homogeneous leptonic synchrotron-self-Compton (SSC) jet models often provide
reasonable descriptions of their overall spectral energy distributions (SEDs). For Cen~A, however, classical one-zone SSC
models (under the proviso of modest Doppler beaming) are unable to satisfactorily account for its core SED up to the highest energies
\cite[cf.][]{chiaberge01,lenain08,abdo10b}. It seems thus well possible, that an additional component contributes to the observed emission
at these energies \cite[e.g.,][]{lenain08,rieger09}. The results presented here indeed provides support for such a consideration.
\begin{figure}[h!]
   \centering
    \plotone{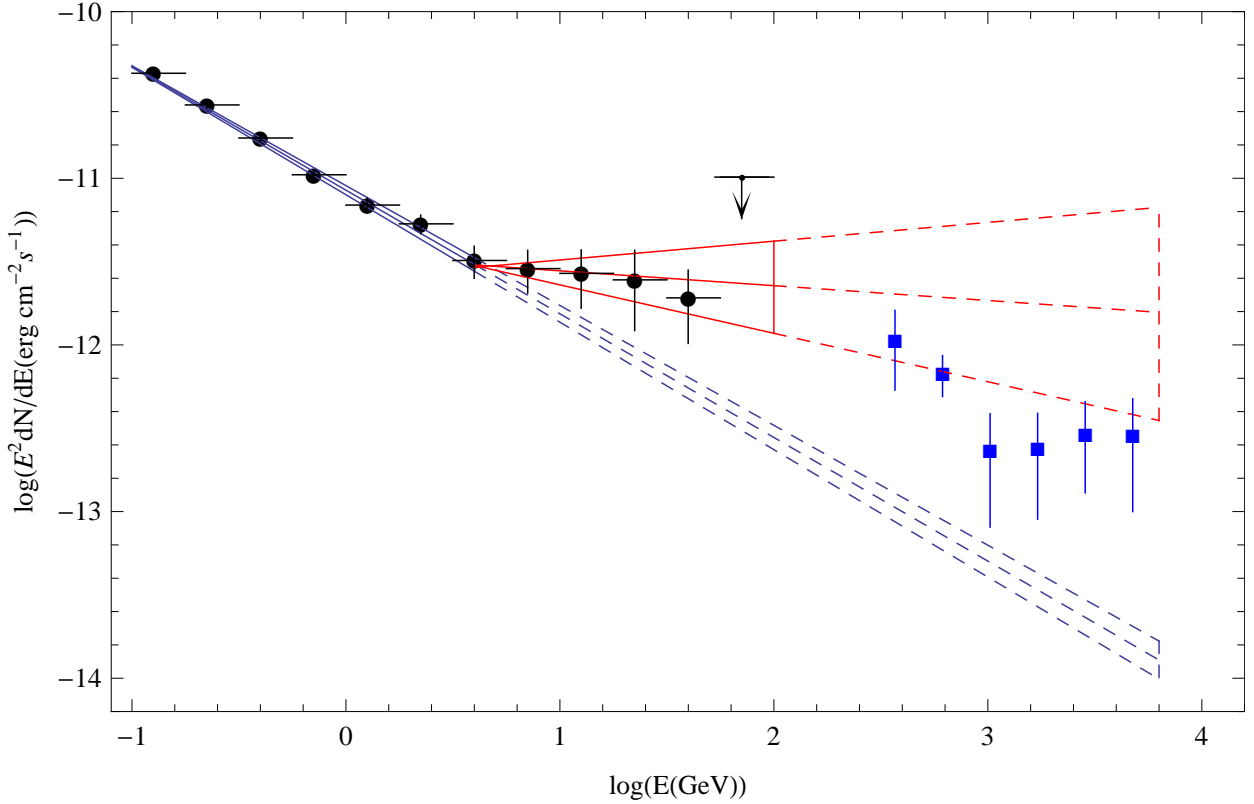}
    \caption{Gamma-ray spectrum for the core of Cen~A from high (\fermi, this work) to very high (H.E.S.S., blue squares) energies. The blue
    bowtie represents a power-law with photon index $2.74$, and the red bowtie a power-law with photon index $2.09$. The dashed lines
    show extrapolations of these models to higher energies. The power-law extrapolation of the low-energy component (blue lines) would
    under-predict the fluxes observed at TeV energies.}
    \label{fg14}
\end{figure}
Our analysis of the 4~yr-data set reveals that the HE core spectrum of Cen~A shows a "break" with photon index changing from $\simeq 2.7$
to $\simeq 2.1$ at an energy of $E_b\simeq4$ GeV. This break is unusual in that the spectrum gets harder instead of softer, while typically
the opposite occurs. For a distance of $3.8$ Mpc, the detected photon flux $F_{\gamma}=(1.68\pm0.04)\times 10^{-7}$ photon cm$^{-2}$s$^{-1}$
for the component below 4 GeV corresponds to an apparent (isotropic) $\gamma$-ray luminosity of $L_{\gamma}(0.1-4~\mathrm{GeV})\simeq
10^{41}$ erg s$^{-1}$. The component above 4 GeV, on the other hand, is characterized by an isotropic HE luminosity of $L_{\gamma}(>4
\mathrm{GeV})\simeq1.4 \times10^{40}$ erg s$^{-1}$. This is an order of magnitude less when compared with the first component, but still larger
than the VHE luminosity reported by H.E.S.S. $L_{\gamma}(>\:250 \:\mathrm{GeV})=2.6\times10^{39}$ erg s$^{-1}$ \cite{aharonian09}.
All luminosities are below the Eddington luminosity corresponding to the black hole mass in Cen~A; nevertheless, they are still quite
impressive when compared with the other nearby radio galaxy M87 containing a much more massive black hole.

Figure~\ref{fg14} shows the gamma-ray spectrum for the core of Cen~A up to TeV energies. As one can see, the flux expected based on
a power-law extrapolation of the low-energy component (below the break) clearly falls below the TeV flux reported by H.E.S.S.. Although
the uncertainties in the photon index are large, it is clear that the spectrum becomes harder above 4 GeV. Remarkably, a simple
extrapolation of the second (above the break) high-energy component to TeV energies could potentially allow one to match the average
H.E.S.S. spectrum. These spectral considerations support the conclusion that we may actually be dealing with two (or perhaps even more)
components contributing to the HE gamma-ray core spectrum of Cen~A. Our analysis of the HE light curves provides some weak indication
for a possible variability on 45 day time scale, but the statistics are not sufficient to draw clear inferences.

The limited angular resolution ($\sim 5$ kpc) and the lack of significant variability introduces substantial uncertainties as to the production
site of the HE gamma-ray emission. In principle, the hard HE component could originate from both a very compact (sub-pc) and/or extended
(multi-kpc) region(s). The double-peaked nuclear SED of Cen~A has been reasonably well-modeled up to a few GeV in terms of SSC
processes occurring in its inner jet \cite[e.g.,][]{chiaberge01,abdo10b}. In this context, the hardening on the HE spectrum above 4 GeV would
indeed mark the appearance of a physically different component. This additional component could in principle be related to a number of
different (not mutually exclusive) scenarios, such as (i) non-thermal processes in its black hole magnetosphere \cite{rieger09}, (ii) multiple
SSC-emitting components (i.e., differential beaming)\cite{lenain08} or (iii) photo-meson interactions of protons in the inner jet
\cite{kachelriess10,sahu12}, (iv) $\gamma$-ray induced pair-cascades in a torus-like region (at $\sim 10^3 r_s$) \cite[e.g.][]{roust11} (v)
secondary Compton up-scattering of host galaxy starlight \cite{stawarz06} or (vi) inverse-Compton (IC) processes in the kpc-scale jet
\cite[e.g.][]{hardcastle11}. What concerns the more compact scenarios (i)-(iv) just mentioned: Opacity considerations do not a priori exclude
a near-BH-origin, but could potentially affect the spectrum towards highest energies \cite[e.g.][]{rieger11}. A SSC multi-blob VHE contribution,
on the other hand, requires the soft gamma-rays to be due to synchrotron instead of IC processes, in which case correlated variability might
be expected. Photo-meson ($p\gamma$) interactions with, e.g., UV or IR background photons ($n_{\gamma}$) require the presence of
UHECR protons, which seems feasible for Cen~A. However, as the mean free paths $\lambda \sim 1/(\sigma_{p\gamma} n_{\gamma}
K_p)$ of protons through the relevant photon fields are comparatively large, usually only a modest fraction of the proton energy can be
converted into secondary particles. Models of this type thus tend to need an injection power in high-energy protons exceeding the
average jet power of $\sim10^{43-44}$ erg/s \cite[e.g.][]{yang}.
The efficiency of IC-supported pair cascades in Cen~A, on the other hand, appears constrained by low accretion modes and the possible
absence of a dust torus. Considering the more extended scenarios (v)-(vi): Partial absorption ($\sim 1\%$) of nuclear gamma-rays by
starlight in the inner part of the host galaxy, and subsequent up-scattering of starlight photons could potentially introduce another HE
contribution. However, the efficiency for this process is low, so that a high VHE injection power into the ambient medium is required, and
the predicted spectral shape does not seem to match well. Compton-upscattering of starlight photon by energetic electrons in the kpc-scale
jet also seems to have difficulties in reproducing the noted HE characteristics.

Finally, let us mention that gamma-ray production may perhaps also be related to relativistic protons interacting with the ambient gas in the
large (kpc) scale regions, e.g., the overall elliptical galaxy NGC~5128 or the densest part of its dust lane. Note that the $\gamma$-ray
luminosity $\approx 10^{41} \ \rm erg/s$ above 100 MeV is larger by two orders of magnitude than the $\gamma$-ray luminosity of the
Milky Way, which could be related to a higher rate of cosmic-ray production and a more effective confinement in the case of NGC~5128.
Moreover, gamma-rays might also be produced in a diluted $R_{\rm halo} \sim  30$ kpc (halo) region of this galaxy. Despite the low density
of gas, gamma-ray production on characteristic timescale $t_{\rm pp} \approx 3 \times 10^{9} (n/10^{-2} \rm cm^{-3})^{-1}$~yr can be
effective, even for a relatively fast diffusion of cosmic rays in this region. More specifically, the efficiency could be close to one,
if the diffusion coefficient at multi-GeV energies does not exceed $D \sim R_{\rm halo}^2/t_{\rm pp} \sim  10^{29}$ cm$^2$/s. This seems
an interesting possibility, especially for the second (hard) HE component with photon index close to $2.1$, in the context of its similarity to
the gamma-ray spectrum of the so-called 'Fermi Bubbles' around the center of our Galaxy \cite{su10}. The much higher luminosity (by
$\sim2-3$ orders of magnitude) of the second component compared to the gamma-ray luminosity of the Fermi Bubbles seems quite
natural, given the much larger energy available in Cen~A, in particular in the form of kinetic energy of its jet.

The results presented here provide observational evidence for an additional contribution at the highest energies and a more complex
spectral gamma-ray behavior than previously anticipated. While considerations like those mentioned above may lead one to favor one
production scenario over the other, none of them cannot be easily discarded. In fact, it is well conceivable that several of them contribute
to the observed gamma-ray emission. Definite progress in this regard could be achieved in case of a significant detection of gamma-ray
time variability.




\clearpage


\end{document}